\documentclass[10pt]{article}
\usepackage{epsfig,graphicx}
\usepackage{times,ch2005}

\def\beq{\begin{equation}}
\def\eeq#1{\label{#1}\end{equation}}
\def\eeqn{\end{equation}}

\def\beqa{\begin{eqnarray}}
\def\eeqa#1{\label{#1}\end{eqnarray}}
\def\eeqan{\end{eqnarray}}

\def\Dslash{\not{\hbox{\kern-4pt $D$}}}
\def\dslash{\not{\hbox{\kern-2pt $\del$}}}

\newcommand{\tev}{\ensuremath{\mathrm{\,Te\kern -0.1em V}}\xspace}
\newcommand{\gev}{\ensuremath{\mathrm{\,Ge\kern -0.1em V}}\xspace}
\newcommand{\mev}{\ensuremath{\mathrm{\,Me\kern -0.1em V}}\xspace}
\newcommand{\kev}{\ensuremath{\mathrm{\,ke\kern -0.1em V}}\xspace}
\newcommand{\ev}{\ensuremath{\mathrm{\,e\kern -0.1em V}}\xspace}
\newcommand{\gevc}{\ensuremath{{\mathrm{\,Ge\kern -0.1em V\!/}c}}\xspace}
\newcommand{\mevc}{\ensuremath{{\mathrm{\,Me\kern -0.1em V\!/}c}}\xspace}
\newcommand{\gevcc}{\ensuremath{{\mathrm{\,Ge\kern -0.1em V\!/}c^2}}\xspace}
\newcommand{\mevcc}{\ensuremath{{\mathrm{\,Me\kern -0.1em V\!/}c^2}}\xspace}

\def\mus  {\ensuremath{\rm \,\mus}\xspace}

\def\mus        {\ensuremath{\,\mu{\rm s}}\xspace}    

\begin{document}


\Title{High Energy All Sky Transient Radiation Observatory}
\bigskip


%
\label{VassilievStart}

%
\author{ V. Vassiliev, S. Fegan \index{Vassiliev, V.} \index{Fegan, S.} }

%
\address{University of California Los Angeles\\
Department of Physics and Astronomy \\
430 Portola Plaza, Box 951547 \\
Los Angeles, CA 90095-1547
}

\makeauthor\abstracts{
The scientific discoveries made by H.E.S.S. during its first year of
operation encourage a reexamination of the open problems in high
energy astrophysics and of the capabilities of the atmospheric
Cherenkov technique, which could be employed to address them. We
report on an initial Monte Carlo exploration of a ground-based
instrument for observing cosmologically distant high energy transient
phenomena. Such observations would require combining an order of
magnitude increase in collecting area over existing instruments, with
a similarly large increase in field of view for routine sky
monitoring, and a factor of five decrease in energy threshold to
extend the visibility range to redshifts exceeding one. A large array
of moderately sized Imaging Atmospheric Cherenkov Telescopes (IACTs)
through which the total array collecting area, array field of view and
array data rates can be distributed, appears to be able to meet these
requirements. To be practically feasible for construction, however, a
cost effective imaging solution, with resolution of $\sim 1$ minute of
arc within $\sim 15$ degrees field of view needs to be found. If image
intensifier based technology provides such a solution in future, then,
combined with super-fast parallel data processing, it may become a
technological foundation for the next breakthrough in ground-based
$\gamma$-ray astronomy Ref.~\cite{JelleyPorter}.}

\section{Introduction}

The construction of a ground-based $\gamma$-ray observatory in the
post-GLAST epoch will be motivated by many astrophysical goals. In the
$1-200$ GeV energy domain several hundred low flux GLAST sources will
require follow up identification and spectral measurement. The
exploration of the $\gamma$-ray sky with sensitivity better than that
which can be achieved by GLAST ($10^{-10}$ cm$^{-2}$ s$^{-1}$,
5$\sigma$) and the study of low fluence, very high energy transient
phenomena in the Universe is very compelling. In the $1-200$ GeV
energy domain, study of the population and cosmological evolution of
rapidly flaring AGN (and possibly VHE GRBs) is possible, due to the
rapid decline of the $\gamma$-ray opacity of the Universe. If VHE
radiation from AGNs is emitted from within a few Schwarzschild radii
of the central black holes, as suggested by the rapid variability of
these objects, then the possibility of probing spatial scales
equivalent to a fraction of a microsecond of arc through the
interpretation of resolved VHE light curves may provide an attractive
opportunity to research matter under conditions of strong gravity. The
diffuse cosmological radiation and its evolution in UV, visible, near-
and mid-IR can also be explored utilizing the pair-production
mechanism responsible for the opacity of the Universe. The very high
energy properties of space-time, large scale magnetic fields affecting
the cascading of photon beams in intergalactic space, and transient
phenomena of more exotic natures, such as the evaporation of black
holes or annihilation of cosmological defects represent other
opportunities for investigation in this energy regime. Astrophysical
observations in the energy range above $200$ GeV with sensitivities
5-10 times better then that of currently existing observatories would
enable deep exploration of galactic objects, which has been proved
fruitful by recent discoveries made by the H.E.S.S.\ observatory
Ref.~\cite{HESSskysurvey}. Motivated by these scientific opportunities
we explore the potential of a large array of mid-size IACTs to conduct
sky monitoring with wide field of view and moderate collecting area or
pointed observations with collecting areas exceeding $1$ km$^2$.

\section{Basic IACT array design considerations}

\begin{figure}[ht]
\resizebox{\textwidth}{!}{
\resizebox{!}{\textheight}{\includegraphics{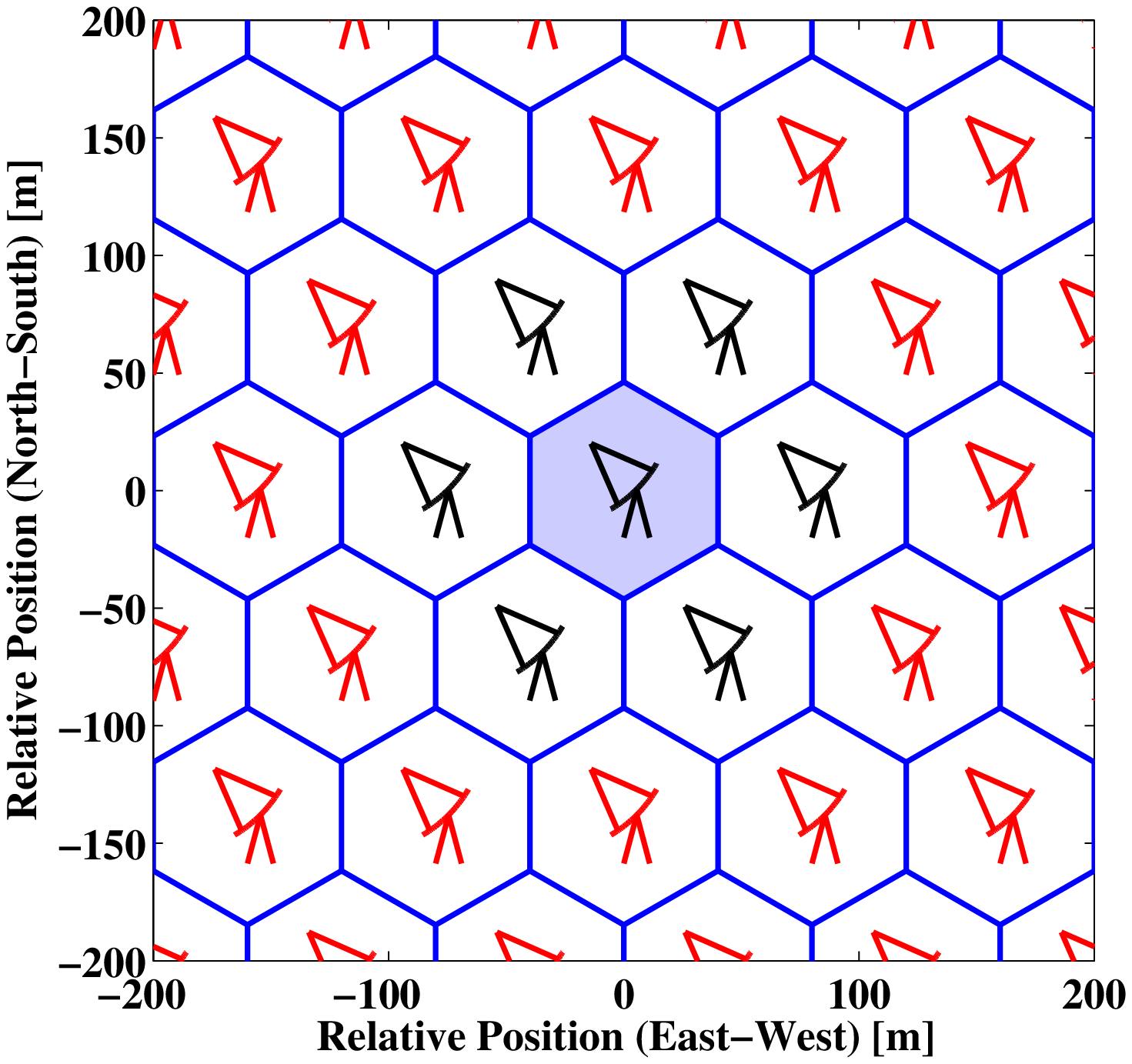}}
\resizebox{!}{\textheight}{\includegraphics{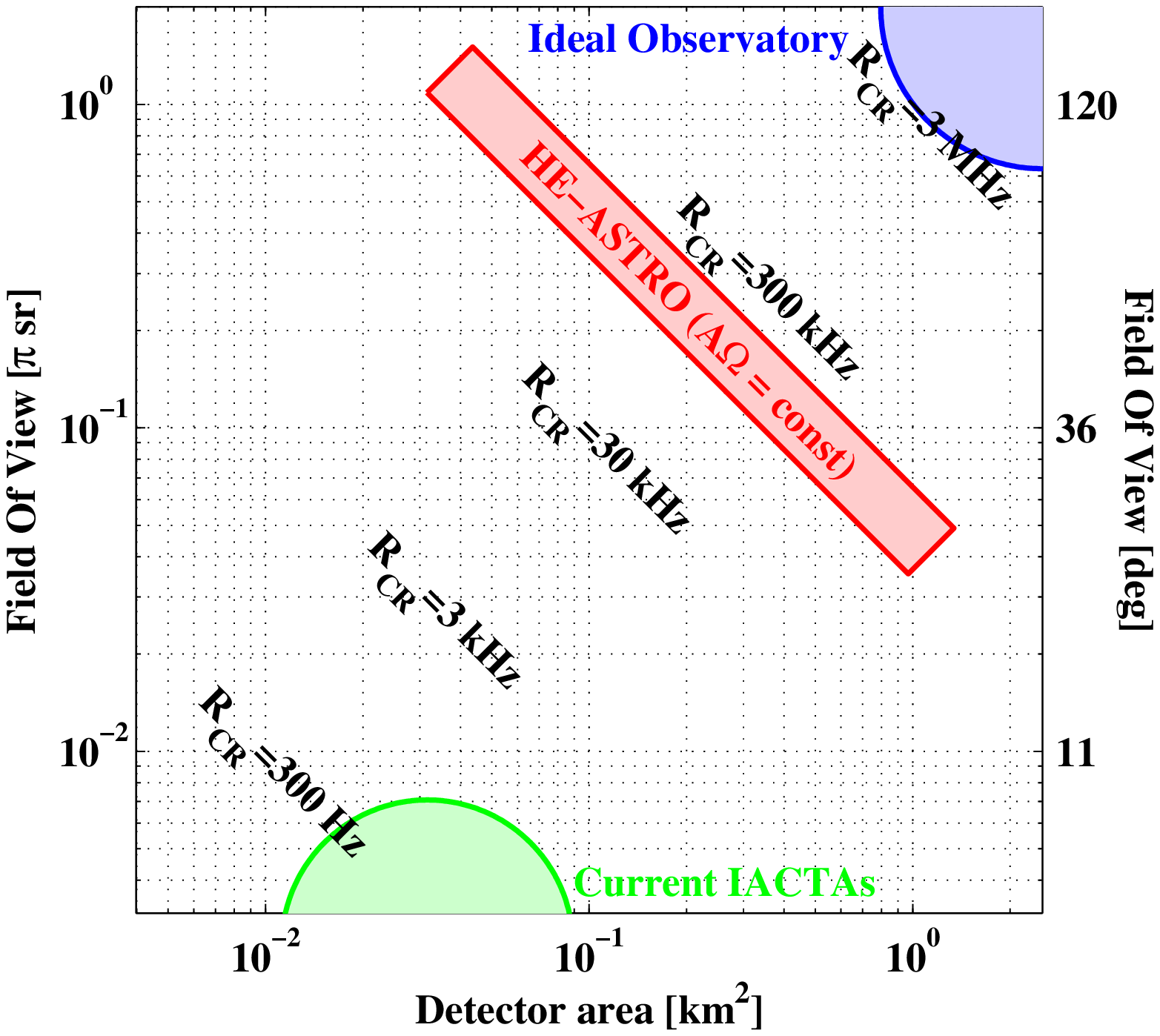}}}
\caption{Left: Illustration of telescope array concept. The separation 
between telescopes is $80$ m, giving the area of a single cell area
(shaded) as $5.5\times10{^3}$ m${^2}$. The original VERITAS layout
is shown with the black color telescopes. Right: Field of View versus
Collecting Area phase space. Areas occupied by present day IACT
arrays, HE-ASTRO and ``Ideal'' observatories are shown. The rough
estimate of detection rates of cosmic rays is indicated as isolines
of constant etendue.}
\label{fig:Fig1}
\end{figure}

To achieve a resolution of AGN flaring on a few minute time scale from
a ``Markarian-421-like'' object placed at a distance of $z=1$, a
detector collecting area of order $1$ km$^2$ or larger is
required. This is a factor of one hundred larger than the Cherenkov
light pool area produced by $\gamma$-rays with $E<100$ GeV, which
determines the effective aperture of the present day ground based
observatories at low energies.  Therefore an array consisting of a few
hundred telescopes may provide a starting point for an instrument
compatible with this requirement. The large collecting area of the
atmospheric Cherenkov technique, as compared to the physical detector
size, has been appreciated for many years. It is evident, though, that
this conclusion depends on the energy of the detected $\gamma$-ray
events and it eventually fails at low energies. For example, the
original design of the VERITAS array, seven telescopes in a hexagonal
arrangement with $80$ m separation, has a $\gamma$-ray collecting area
of $\sim 5 \times 10^{3}$ m$^2$ at energies around $50$ GeV, equal to
the area of a hexagonal cell occupied by a single telescope in the
array (see Figure~\ref{fig:Fig1} for illustration).  The concept of an
IACT ``cell'', first introduced in
Ref.~\cite{IACTarraysI,IACTarraysII}, provides a convenient framework
for evaluating the performance of large, but finite, IACT arrays at
low energies when the effective collecting area per cell is comparable
or smaller than the cell size.

In the low energy regime the collecting area of a large IACT array
simply scales with the number of cells in the array. At high energies,
however, the scaling law is violated when an increasing number of
triggering $\gamma$-rays occur outside of the array and the full
collecting area becomes larger than its physical size. This occurs at
energies of a few TeV for a $1$ km$^2$ array.  In the high energy
regime the collecting area grows roughly in proportion to the
perimeter of the array and therefore as the square root of the number
of cells (or telescopes) in it. Because of the different scaling of
collecting area at low and high $\gamma$-ray energies, a large array
of IACTs, spaced at a distance sufficient for accurate stereoscopic
reconstruction, naturally has significantly improved low energy
performance. Increasing the number of telescopes in the array to much
larger than a few, shifts the peak of the differential detection rate
of $\gamma$-rays towards lower energies. This approach is
complementary to increasing the aperture of individual telescopes or
building IACT arrays at high elevations in providing for operation in
the $20-200$ GeV domain. However, unlike these alternatives it also
provides a $\sim 1$ km$^2$ collecting area. The exploration of this
potential is one of the main goals of this submission.

In addition to providing a large collecting area for pointed
observations and improved collecting area at low energies, a large
array of telescopes facilitates operation in ``sky survey mode'', in
which individual telescopes are uniformly spread over $\sim$2 sr of
the sky. In this mode the collecting area at low energies is
comparable to that of the VERITAS or H.E.S.S. instruments. For
stereoscopic reconstruction, events must be seen by at least three
telescopes. This coupling between telescopes in sky survey mode
requires that the field of view of individual instruments be $>15$
degrees. Large solid angle sky coverage through the combination of
moderately large field of view telescopes enables highly flexible
operation. Figure~\ref{fig:Fig1} illustrates that such an array allows
for a continuum of operation modes which trade effective collecting area on any
point in the sky for total solid angle coverage, roughly characterized
by the constant etendue condition ($A\Omega=$const).

\begin{table}[htbp]
\begin{center}
\begin{tabular}{|l|l|}\hline
Array composition                 & 217 telescopes on hexagonal grid\\
Elevation                         & 3.5\,km \\
Telescope coupling distance       & 80\,m \\
Internal area                     & $\sim$1.0\,km$^2$ \\
Single Telescope Field of View    & $\sim$15$^\circ$ \\
Single Telescope FoV area         & $\sim$177\,deg$^2$ \\
Reflector Diameter                & $\sim$7\,m \\
Reflector Area                    & $\sim$40\,m$^2$ \\
Quantum Efficiency                & 50\% (200-400\,nm) \\
Mirror Reflectivity               & 64\% (double reflection $0.8 \times 0.8$) \\
Trigger Sensor Size               & $\sim$31.2\,cm \\
Night Sky Background (NSB) rate   & $\sim$3.2\,pe/trigger pixel/20\,ns \\
Single Telescope NSB Trigger Rate & 1\,KHz \\
Trigger Sensor Pixel Threshold    & 21\,pe (DC) \\
Single Telescope CR trigger rate  & $\sim$30\,kHz \\
Energy Range                      & 20\,GeV\,--\,200\,TeV \\
Differential Detection Rate Peak  & $\sim$30\,GeV ($\gamma$-rays) \\
Image Sensor Pixel Size           & 0.0146$^\circ$ \\
Readout Image                     & 128\,$\times$\,128 pixels \\
Readout Image Size                & 1.875$^\circ$ $\times$ 1.875$^\circ$ \\
NSB Rate per Image Sensor Pixel   & 0.032 (per 20 nsec gate) \\
ADC (Image Sensor)                & 8\,bit (S/N improved) \\
Image Integration Time            & 20\,ns \\
Data Acquisition Rates             & $\sim$80\,Mb/sec/node \\\hline
\end{tabular}
\caption{HE-ASTRO Specifications for simulations} 
\label{tab:HEASTROparameters}
\end{center}
\end{table}

The large collecting area and solid angle coverage gained by large
arrays of IACTs is made possible by the combined contribution of
individual telescopes, each of which contributes only a small portion
to the array collecting area and array solid angle.  This
``distributed operation'' is particularly critical to sustain the high
data rates from cosmic rays in the field of
view. Figure~\ref{fig:Fig1} (right) shows that the design of a single
instrument with equivalent etendue value operating in $\sim 40$ GeV
energy domain would require data processing at the rate of $\sim 300$
kHz or higher. The distributed operation, which is a unique feature
of a large array, makes it possible for a single telescope to produce
data at a rate ten times lower, $\sim 30$ kHz, which is manageable
even with present day technology.

It appears that a large array of mid-size IACTs can successfully
attain the requirements of large collecting area ($>1$ km${^2}$),
large field of view ($\sim 2$ sr), and reasonable data rate per
telescope ($\sim 30$ kHz) imposed by the scientific goal of
observation of VHE transient phenomena in the $\sim 40$ GeV domain.
There are two questions which must be seriously addressed in studying
the feasibility of such an instrument: its capabilities in the
required energy regime and the technical problem of finding
cost-effective solutions for its design. In this paper we report on
detailed simulation studies of the low energy performance of the
array. We do not attempt to address the cost issues, though some
thoughts were reported in the presentation to this meeting
Ref.~\cite{ch2005presentation}.

\section{Simulations}

The simulation studies were performed using CORSIKA 6.2
Ref.~\cite{CORSIKA}, assuming idealized optics with no
aberrations. Table~\ref{tab:HEASTROparameters} summarizes the
simulation input parameters. Although we simulated $7$ m aperture
telescopes with $50$\% quantum efficiency, the results are equally
applicable to $10$ m telescopes with $25$\% quantum efficiency.  The
focal plane instrument has two light sensors: the ``trigger sensor''
and ``image sensor''.  The size of the pixels in each sensor was
optimized independently to achieve the best performance at low
energies. 

\subsection{Trigger optimization}

\begin{figure}[ht]
\resizebox{\textwidth}{!}{
\resizebox{!}{\textheight}{\includegraphics{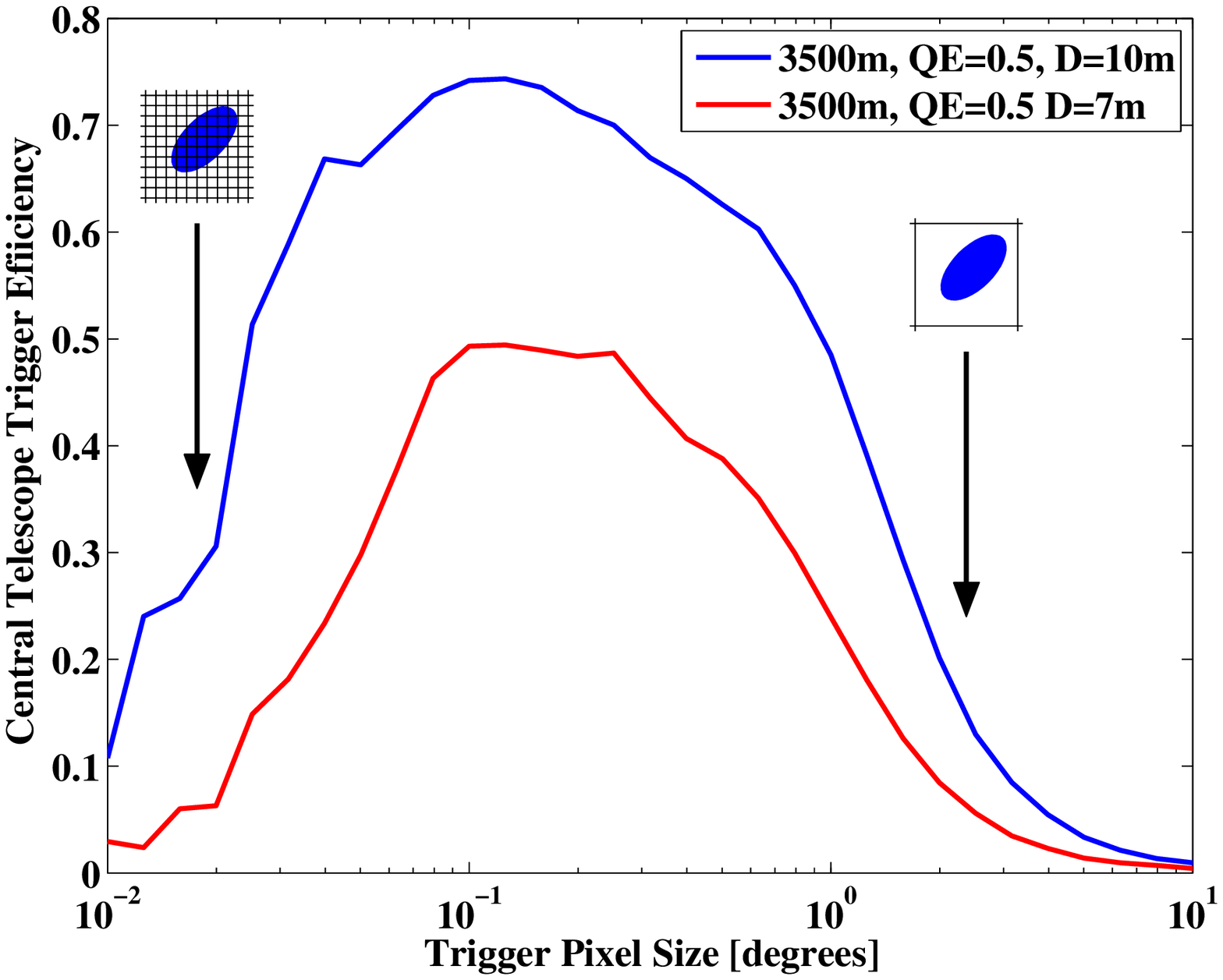}}
\resizebox{!}{\textheight}{\includegraphics{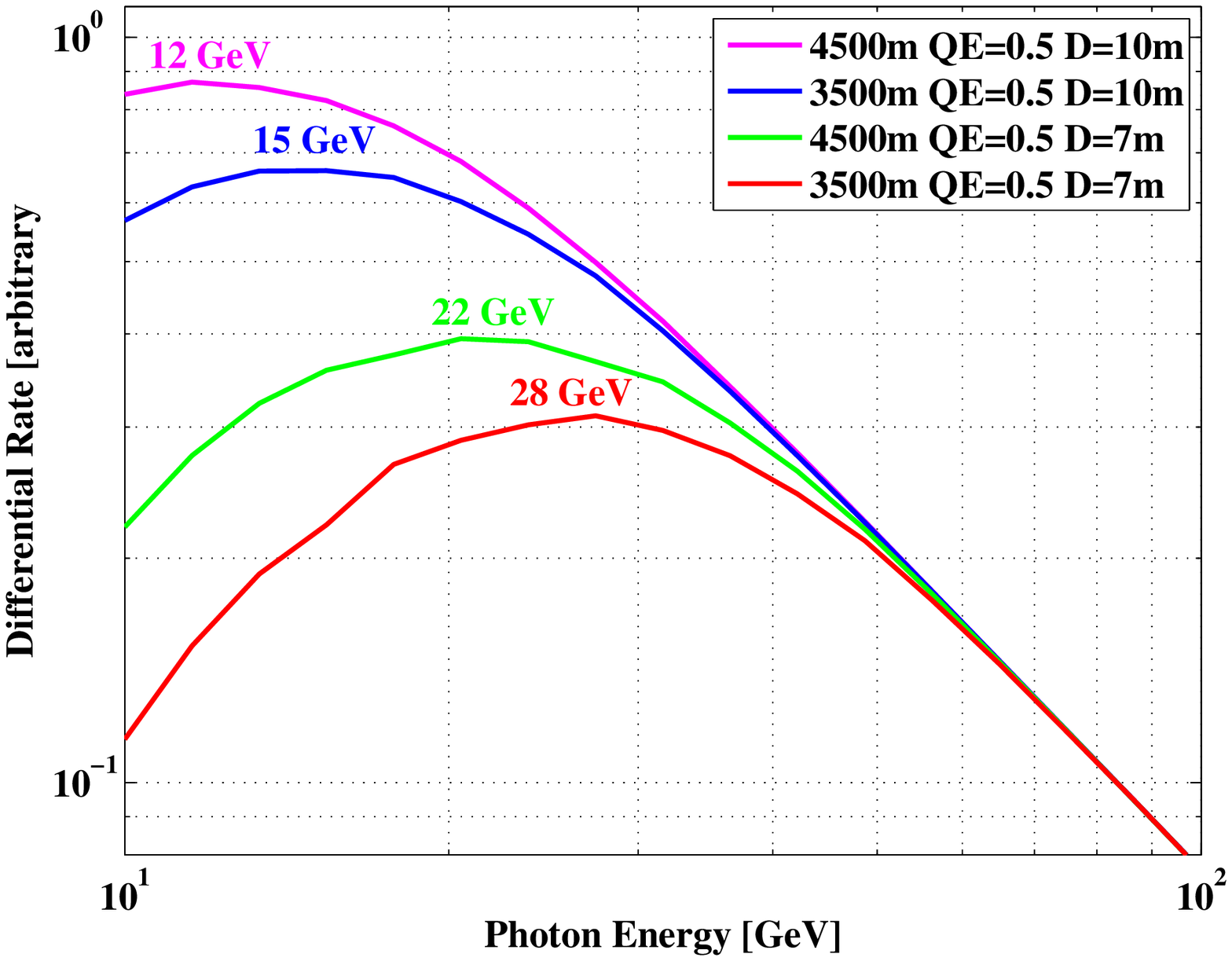}}}
\caption{Left: Trigger efficiency as a function of trigger sensor pixel 
size for a $3.5$ km elevation array with detector quantum efficiency
of $50$\%. The two curves show different telescope apertures: $7$ and
$10$ m. Right: Differential detection rates of $\gamma$-rays from a
source with a power law spectrum with spectral index of $-2.5$. The
effect of elevation and telescope aperture on the position of the peak
is shown.}
\label{fig:Fig2}
\end{figure}

The dependence of the $\gamma$-ray detection efficiency on the trigger
sensor pixel size can be roughly understood as follows.  When the
pixel size is much larger than the characteristic scale in the shower
image the signal remains constant while night sky noise scales with
the pixel area. Thus, the ratio of signal to square root of noise
declines inversely proportional to the size of the pixel. In the
opposite limit, subdividing the Cherenkov image over many pixels, each
much smaller than the characteristic scale of the image, reduces the
signal in proportion to the pixel area leading to a decline in the
ratio of signal to square root noise in proportion to pixel
size. These trends are confirmed by simulations and shown in
Figure~\ref{fig:Fig2}. It was found that the optimum pixel size for
efficient triggering of the telescope located at the center of a cell
is in the range $0.05 - 0.25$ degrees.  The optimum value is
relatively insensitive to telescope aperture, elevation, and even
energy of the $\gamma$-ray. Individual telescopes were triggered when
a single pixel of the trigger sensor exceeded a given threshold, set
by the requirement of a 1kHz rate of night sky background fluctuations
in the 15$^\circ$ field of view. For the parameters shown in
Table~\ref{tab:HEASTROparameters} this threshold was calculated to be
$21$ pes.

The current generation of IACTs use a topological trigger which
imposes a requirement that two, three or four adjacent PMT channels
exceed a certain threshold, e.g. $\sim 5.6$ pe in each of three
neighboring pixels in the case of the VERITAS array. The collection of
these adjacent pixels forms an effective trigger sensor pixel size,
which in the case of VERITAS is larger than the optimum found in our
studies. In VERITAS and H.E.S.S. the same light sensor (an array of
PMTs) is used to image the cascade and trigger the telescopes.
Topological triggers are necessary to compromise between the different
resolution scales required for accurate event reconstruction and
identification (imaging) and efficient triggering when a single light
sensor implements both functions simultaneously. In the next section
we discuss the optimum image sensor pixel size.

Another benefit of topological triggers which is not often discussed
is their ability to filter Cherenkov flashes produced by the
overwhelming background of distant single particles which would
otherwise trigger each telescope in an array.  There is enough light
generated by a single particle, e.g. muon, with impact parameter as
large as the radius of the Cherenkov light pool $\sim 120$m to trigger
all telescopes of the VERITAS or H.E.S.S. array and therefore produce
an array trigger. For large impact parameters ($>\sim 25$m), the image
of a single particle is normally contained within a single pixel of 
the camera and therefore the topological trigger inhibits the
telescope trigger. Since the particle is local to at most one
telescope, an array trigger cannot be generated by single particles.
In practice, non-idealities of the optics may spill the image into
two or three fine pixels in a high resolution imaging camera and
therefore generate an array trigger. Large aperture IACT telescopes
with fine pixels may be particularly sensitive to this effect.

Simulations of the array trigger as a function of trigger sensor pixel
size indicate that the optimum is in the range, from $\sim0.05$ to
$0.13$ degrees with the peak at $0.08$ degrees for all $\gamma$-ray
photons in the $10$ to $100$ GeV interval. For our simulations we used
a trigger pixel size of $0.15$ degrees, the choice arising from
considerations of potential cost. Figure~\ref{fig:Fig2} shows the
differential detection rate of $\gamma$-ray photons. The curves shown
demonstrate that although there is some reduction of peak energy with
increasing elevation, the effect of higher quantum efficiency or
larger telescope aperture dominates. Construction and operation costs
at higher elevation might not be justified when compared with the
potentially lower cost of investment into improvement of the
telescopes and/or photon detectors. Thus we used elevation of $3.5$ km
for the remainder of our study.

\subsection{Imaging optimization}

In this study we attempted to understand the degree to which
reconstruction of the arrival direction and identification of
$\gamma$-rays could be improved through optimal image pixellation.
When choosing the size of the image pixels, the effects of degradation
due to the optical system and the scales inherent to the shower
physics should be considered. The degree of image distortion due to
the optical system increases with telescope aperture. For example,
image blurring due to the depth of field effects, caused by the finite
distance to the source of light, produces defocusing on the order of
$10$ minutes of arc for a $30$ m telescope imaging a point source at
at distance of $~\sim 10$ km. Comatic off axis aberrations for
practical, moderate focal length designs might be of similar or larger
order. The non-zero size of the primary mirror also limits the
accuracy to which the impact position of the Cherenkov photons can be
known. This loss of information degrades the accuracy to which the
shower core position and arrival direction can be reconstructed. This
effect is complex but seems to become important for dim images when
improvements through statistical averaging is not possible.
In view of these considerations we chose to perform simulations with
relatively small, $7$ m aperture, ``ideal'' telescopes, by assuming
that the arrival direction of photon can be determined exactly. The
impact point of the photon on the reflector surface is, however, lost.

\begin{figure}[ht]
\resizebox{\textwidth}{!}{
\resizebox{!}{\textheight}{\includegraphics{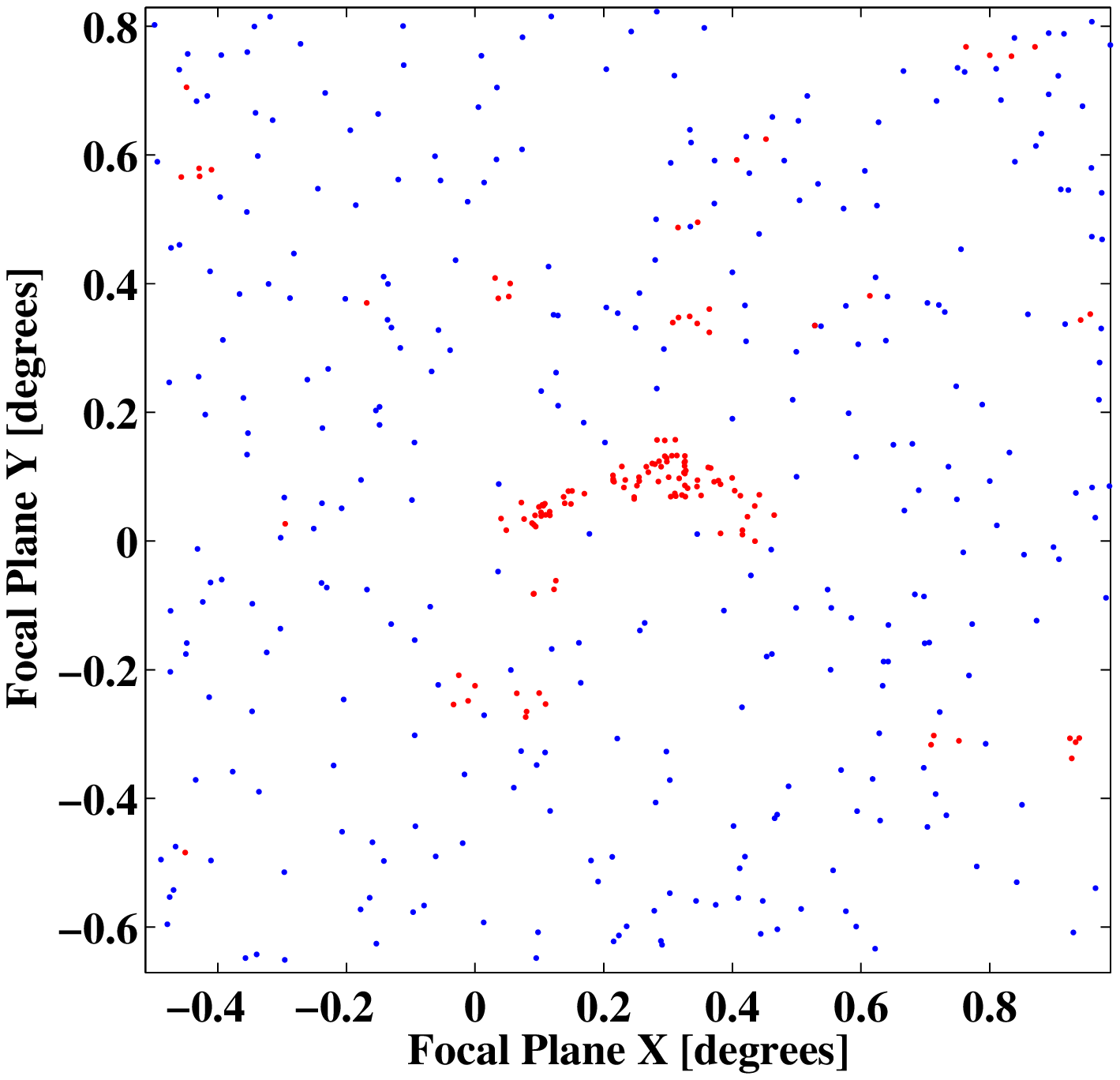}}
\resizebox{!}{\textheight}{\includegraphics{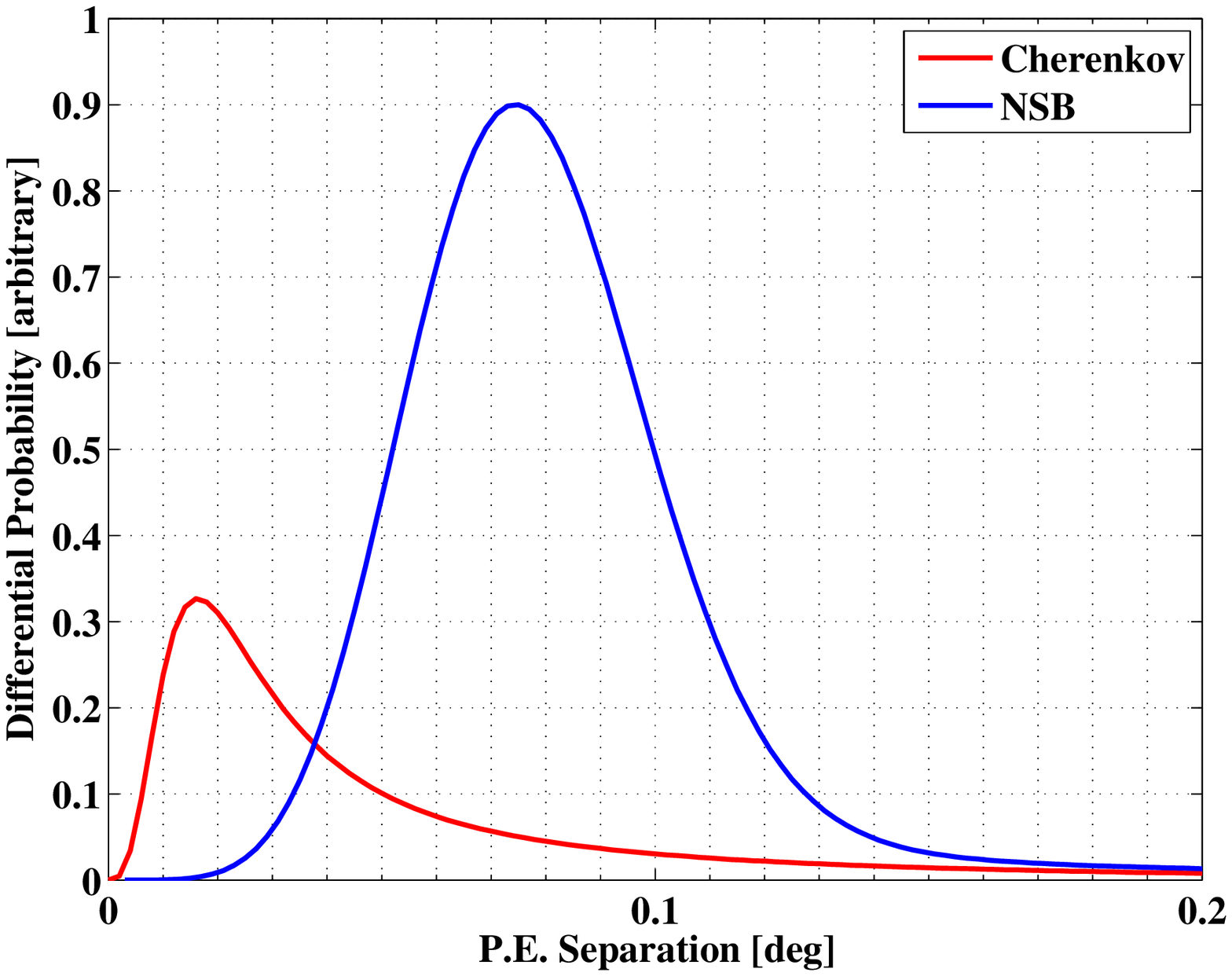}}}
\caption{Left: A typical $42$ GeV event. The Cherenkov photo-electrons 
from the atmospheric cascade are shown in red. The core of the image
is clearly seen as a group of photo-electrons with average separation
of $\sim 0.015$ degrees. Right: Differential probability of
photo-electron to photo-electron angular distance for Cherenkov (red)
and NSB photo-electrons (blue). The plot indicates the distribution of
characteristic radius of the Voronoi cell occupied by each photon. For
this size telescope the peaks are separated.}
\label{fig:Fig3}
\end{figure}

A typical $40$ GeV event as seen by an ideal telescope is shown in
Figure~\ref{fig:Fig3}. There are three distinct angular scales in this
image. The first is determined by the area occupied by a single
photo-electron of night sky background. It can be calculated by
averaging the square root of the area of the Voronoi cells, i.e. those
regions of the focal plane closer to each NSB photo-electron than to
any other. Figure~\ref{fig:Fig3} (right) indicates that for the chosen
telescope aperture and the night sky background noise rate this scale
is $~0.075$ degrees, given by the peak of the distribution. The second
scale is determined by the very dense distribution of photo-electrons
induced by the photons emitted from the innermost core of the
atmospheric cascade. The particles radiating these Cherenkov photons
have not been deflected by multiple Coulomb scattering, due to
their relatively high energy ($E>\sim 100$ MeV), and most accurately
trace the arrival direction of the primary $\gamma$-ray. The extraction
of these photons from the cascade image by requiring very small
Voronoi cell size or, equivalently, a high number of photo-electrons
over a small area, should improve angular resolution. Our simulations
confirm that the angular resolution of reconstructed arrival direction
does improve with finer pixellation of the camera until the
typical angular scale determined by the transverse size of the shower
core image is reached. This size is approximately a few minutes of
arc, as the core of the highest energy particles in the cascade has an
extent of just a few meters, Ref.~\cite{CHESS}, and the typical
observing distance of sub-100 GeV cascades is on the order of $10$ km.
Similar findings were also reported by Prof.~Hofmann in his submission
to this conference Ref.~\cite{Hofmann-ch2005}. Finally, the third
angular scale in the shower image is that of the ``fuzzy'' structures
surrounding the cascade core which generate the extended tail in the
distribution shown in Figure~\ref{fig:Fig3}. These photo-electrons,
produced by relatively low energy particles (less than $\sim 100$ MeV
but larger than $20$ MeV), are coming from distances as large as
$\sim60$ m from the shower core. The differences in the distribution
of these photo-electrons in $\gamma$-ray and hadronic induced cascades
have been effectively used by practitioners of the IAC technique
through the implementation of various cuts, generically called
``shape'' cuts.

Two conclusions follow from our study. First, it is beneficial for
angular reconstruction of the arrival direction of primary
$\gamma$-rays to improve the imaging resolution of IACT cameras to the
$\sim 1$ minute of arc scale, assuming that the telescope optics are
compatible with this requirement. The fine resolution required for the
image sensor is $5--10$ times smaller than the size of the optimum
pixel of the trigger sensor. This suggests that the functionality of
imaging and triggering may require different hardware implementations
in future ground-based $\gamma$-ray instruments. One minute of arc
resolution scale for cascade imaging implies $\sim 10^{6}$ pixels per
camera of 15 degrees field of view. Traditional arrays of PMTs or even
MAPMTs do not seem to provide a practical, cost-effective
solution. A new technology utilizing CMOS, CCD, or perhaps SiPMs and
APDs with multi-million pixel image sensors might be a necessity. The
second conclusion is that multiple image cleaning procedures may
become necessary to get the best performance from high resolution
cameras on future IACTs. A single night sky background cleaning
procedure, as is currently the standard approach in the IAC
technique, may not be simultaneously adequate for angular
reconstruction, background suppression and energy determination, since
the information required for these tasks is contained in different
scales in the image.

\subsection{Performance}

The collecting area and differential detection rate of HE-ASTRO for
$\gamma$-rays for the point source and all sky modes are shown in
Figure~\ref{fig:Fig5}. In point source mode the collecting area is
$\sim 1$ km$^2$ at 30 GeV. At this energy the triggering efficiency
for $\gamma$-rays impacting within the array is $>90\%$ while
$\gamma$-rays impacting outside the array do not contribute much to
the total array collecting area. Below this energy the trigger
efficiency falls, leading to a decline in the overall collecting
area. Above 30 GeV the influence of events impacting outside the array
grows, and the total collecting area reaches $3-4$ km$^2$ at 10 TeV,
substantially larger then the physical size of the array. The large
fields of view of the individual telescopes allows them to trigger on
distant high energy events and the $\sim1.4$ km baseline across the
full array combined with the high resolution image sensor enables
stereoscopic reconstruction. In all sky mode, the detection collecting
area in the 50-200 GeV energy regime is approximately that of a seven
telescope VERITAS array, increasing to $>1$ km$^2$ at higher energies.

\begin{figure}[ht]
\resizebox{\textwidth}{!}{
\resizebox{!}{\textheight}{\includegraphics{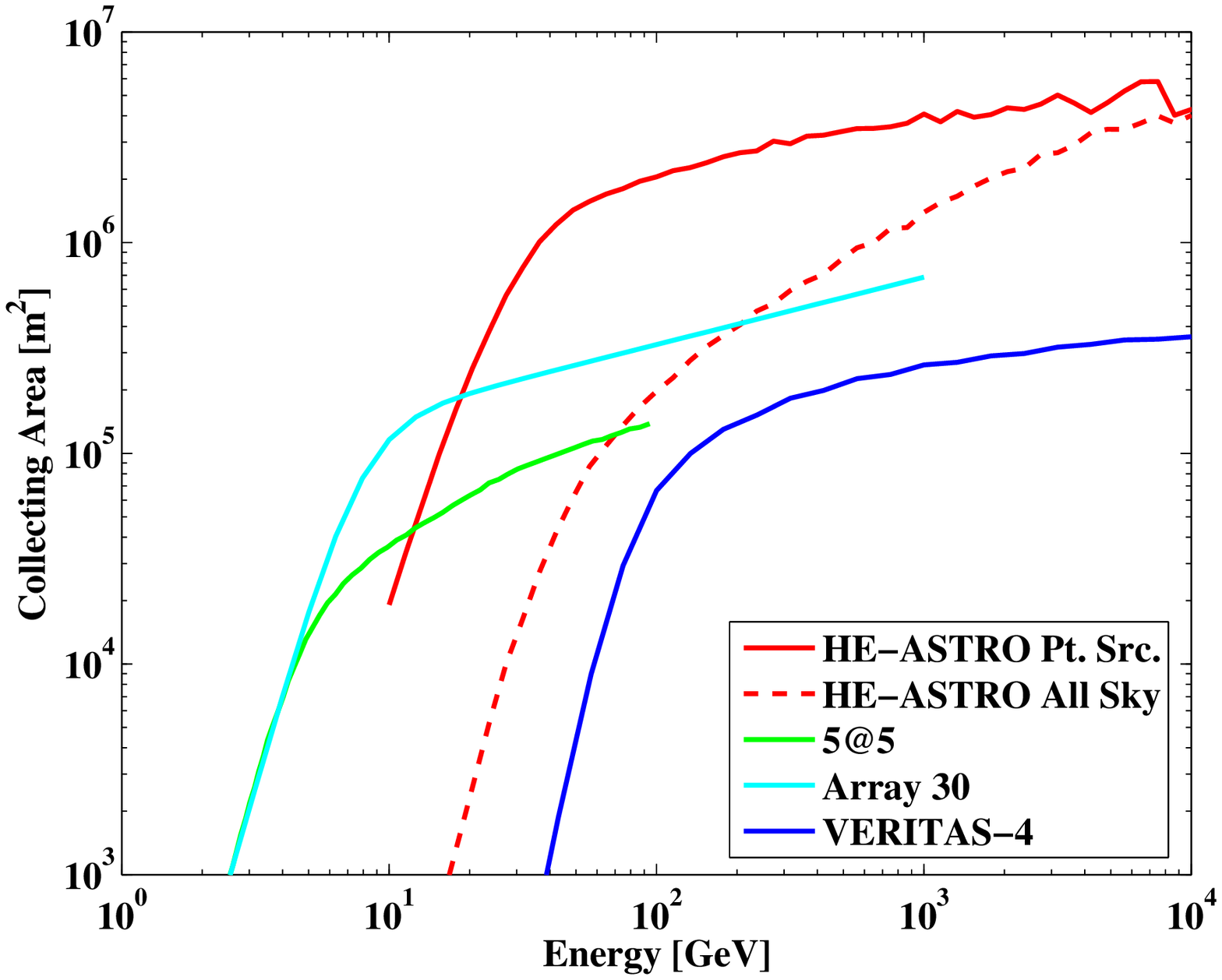}}
\resizebox{!}{\textheight}{\includegraphics{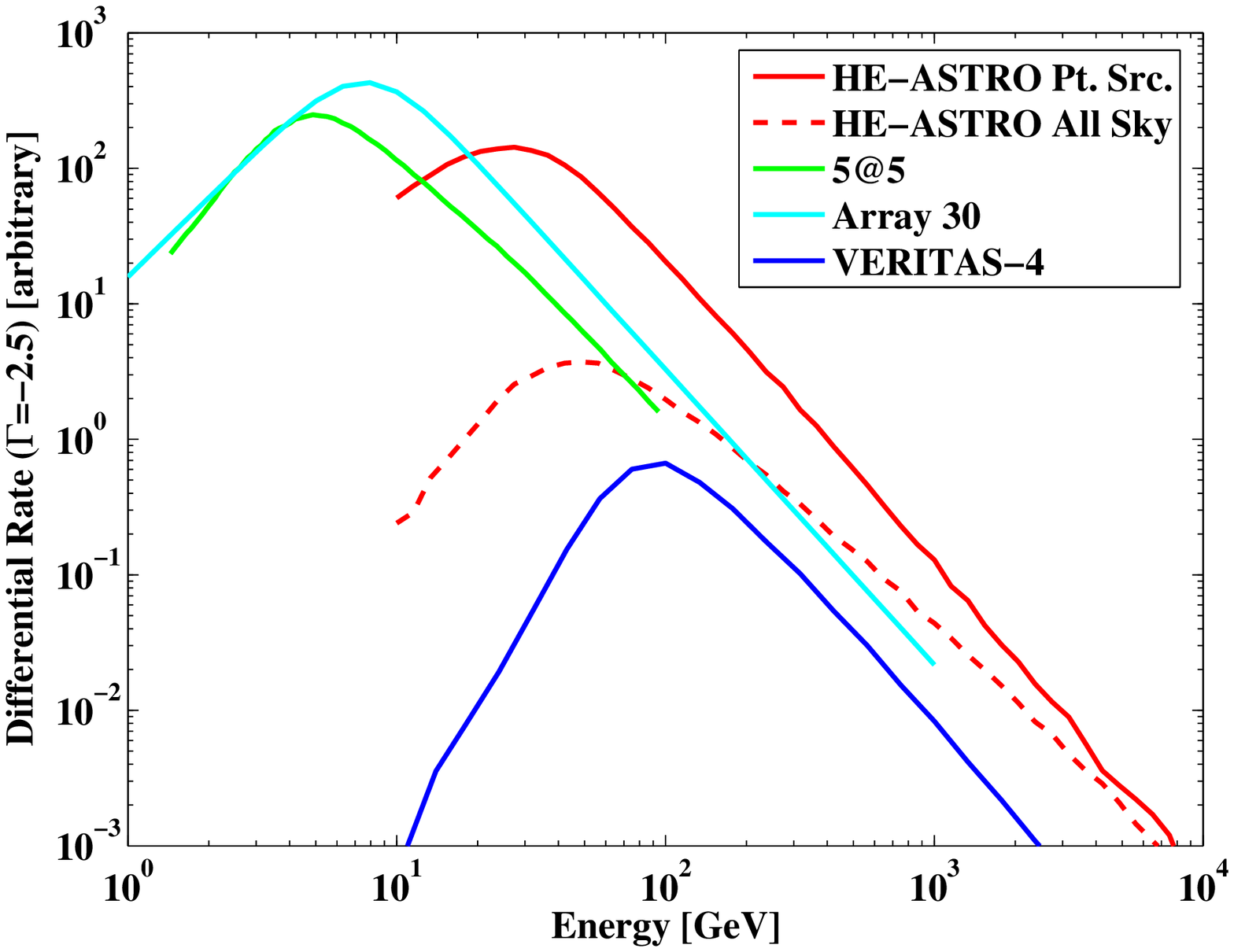}}}
\caption{Effective collecting area (left) and differential detection
rate (right) of HE-ASTRO array in point source mode (solid red) and
all sky mode (dashed red). Shown for comparison are the equivalent
curves for VERITAS-4 (blue), the 5@5 array (Ref~\cite{FiveAtFive}; green)
and the five telescope configuration of the 30 m STEREO ARRAY 
(Ref~\cite{StereoArray30}; cyan).}
\label{fig:Fig4}
\end{figure}

Figure~\ref{fig:Fig5} depicts the flux limit (in mCrab) to which an
all sky instrument with VERITAS-like sensitivity can survey in one
year, assuming that it is located on the equator. From comparison of the
collecting areas, it is expected that the sensitivity of HE-ASTRO in
full sky mode will be at least as good as VERITAS. In fact, improved
background rejection may come from significantly better imaging
resolution, allowing hadronic events to be rejected based on the
difference in the depth of shower maximum in the atmosphere between
proton and $\gamma$-ray induced cascades, thereby improving the
sensitivity.

\begin{figure}[ht]
\resizebox{\textwidth}{!}{
\resizebox{!}{\textheight}{\includegraphics{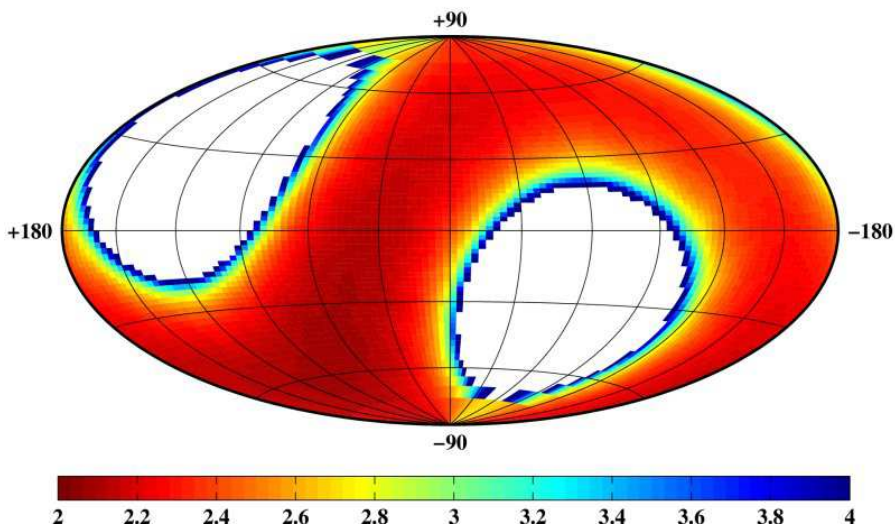}}}
\caption{Flux limit in mCrab to which an all sky instrument with 
VERITAS-like sensitivity could survey in one year, assuming it is 
placed on the equator.}
\label{fig:Fig5}
\end{figure}

\section{Conclusion}

A large array of mid-sized IACTs with moderately large fields of view
may satisfy the technical requirements for observing high energy
transient phenomena throughout the Universe. Such an array will
achieve a factor of one hundred increase in collecting area for
observations of point sources at low energies ($<100$ GeV) over
current instruments. It will be capable of operation in all sky
(survey) mode to monitor $\sim 2$ sr of solid angle with $\gamma$-ray
sensitivity comparable to the VERITAS point source sensitivity. If
built at an elevation of $2.5 - 3.5$ km, the differential $\gamma$-ray
detection rate for pointed observations of a ``Crab-like'' source will
peak in the $30-50$ GeV range or lower, depending on the degree of
improvement in photon detector quantum efficiency achieved over the
coming few years. Each of the telescopes of the array will have a
manageable cosmic ray acquisition rate due to the distributed nature
of operation, in which each telescope contributes only a small part of
the total collecting area and solid angle coverage of the array.
Simulations indicate that the low energy performance of the array
depends on the optimal choice of pixellation for trigger and image
sensors of the telescopes ($\sim 0.1$ and $\sim 0.02$ degrees
respectively). The large difference in these angular scales suggests
that the sensors to perform triggering and imaging functions be
separated in future ground based IACT observatories. This design of
the focal plane instrumentation is different from present day
implementations, in which both functions are performed by a single
array of PMTs, and the telescope trigger is formed by combining
electrical signals from adjacent pixels. The feasibility of
constructing a large array of IACTs depends critically on finding a
cost effective solution for manufacturing individual telescopes. From
this point of view, the implementation of an optimally sized trigger
sensor using a mosaic of MAPMTs appears to be roughly compatible with
current costs. However, the small size required for the image sensor
pixel necessitates the exploration of alternatives, such as CMOS, CCD,
SiPMs, and APD arrays. The small physical size of these devices does
not couple well to the large aperture IACTs which necessarily operate
in a photon limited regime. Before the telescope plate scale can be
changed optically to match the physical size of the sensors,
amplification of the Cherenkov image is required. Image intensifier
technology, which has been viewed as promising for use in ground based
IAC $\gamma$-ray astronomy since its foundation Ref.~\cite{JelleyPorter},
may provide a cost effective resolution to this problem.

\section*{Acknowledgments}

The authors thank Rene Ong, Simon Swordy, and Amanda Weinstein for
discussions and suggestions. We acknowledge the Division of Physical
Science, College of Letters and Science, and the Department of Physics
and Astronomy at UCLA for funding the extensive computation facilities
used for this work.

%
\label{VassilievEnd}
 
\end{document}